\documentstyle[12pt]{article}

\def \di{\partial}

\def \a {\alpha}
\def \b {\beta}

\def \eps {\epsilon}

\def \t  {\tau}

\def \det{{\rm det}}

\def \log {{\rm log}}

\def \ra {\rightarrow}

\def \be {begin {equation}}
\def \ee {end{equation}}
\def\tilde{\widetilde}
\def\bar{\overline}

\def\*{\star}
\def\({\left(}		\def\BL{\Bigr(}
\def\){\right)}		\def\BR{\Bigr)}
\def\[{\left[}		
\def\]{\right]}

\def\frac#1#2{{#1 \over #2}}		

\def\half{{1 \over 2}}

\def\2pi{\hbox{$2\pi i$}}

\def\dsl{\raise.15ex\hbox{/}\kern-.57em\partial}
\def\Dsl{\,\raise.15ex\hbox{/}\mkern-.13.5mu D}
\def\th{\theta}		
		\def\Ga{\Gamma}
\def\be{\beta}

\def\ep{\epsilon}
\def\la{\lambda}	\def\La{\Lambda}
\def\de{\delta}		
		
	\def\Sig{\Sigma}

		\def\CC{{\cal C}}
\def\CD{{\cal D}}		\def\CF{{\cal F}}

\def\CM{{\cal M}}	\def\CN{{\cal N}}	
		
\def\CS{{\cal S}}		
	\def\CW{{\cal W}}	
	
\font\numbers=cmss12
\font\upright=cmu10 scaled\magstep1
\def\stroke{\vrule height8pt width0.4pt depth-0.1pt}
\def\topfleck{\vrule height8pt width0.5pt depth-5.9pt}
\def\botfleck{\vrule height2pt width0.5pt depth0.1pt}
\def\Zmath{\vcenter{\hbox{\numbers\rlap{\rlap{Z}\kern
0.8pt\topfleck}\kern
2.2pt
                   \rlap Z\kern 6pt\botfleck\kern 1pt}}}
\def\Qmath{\vcenter{\hbox{\upright\rlap{\rlap{Q}\kern
                   3.8pt\stroke}\phantom{Q}}}}
\def\Nmath{\vcenter{\hbox{\upright\rlap{I}\kern 1.7pt N}}}
\def\Cmath{\vcenter{\hbox{\upright\rlap{\rlap{C}\kern
                   3.8pt\stroke}\phantom{C}}}}
\def\Rmath{\vcenter{\hbox{\upright\rlap{I}\kern 1.7pt R}}}
\def\Z{\ifmmode\Zmath\else$\Zmath$\fi}
\def\Q{\ifmmode\Qmath\else$\Qmath$\fi}
\def\N{\ifmmode\Nmath\else$\Nmath$\fi}
\def\C{\ifmmode\Cmath\else$\Cmath$\fi}
\def\R{\ifmmode\Rmath\else$\Rmath$\fi}

\def\be{ \begin{eqnarray} }
\def\ee{ \end{eqnarray} }

\def\bea{\begin{eqnarray}}
\def\eea{\end{eqnarray}}
\def\nn{\nonumber}

\def\beq{\begin{equation}}
\def\eeq{\end{equation}}
\def\ba{\beq\new\begin{array}{c}}
\def\ea{\end{array}\eeq}
\def\be{\ba}
\def\ee{\ea}

\def\Tr{{\rm Tr}}

\makeatletter
\newdimen\normalarrayskip         
\newdimen\minarrayskip            
\normalarrayskip\baselineskip
\minarrayskip\jot
\newif\ifold             \oldtrue            \def\new{\oldfalse}
\def\arraymode{\ifold\relax\else\displaystyle\fi}
\def\eqnumphantom{\phantom{(\theequation)}}     
\def\@arrayskip{\ifold\baselineskip\z@\lineskip\z@
     \else
     \baselineskip\minarrayskip\lineskip2\minarrayskip\fi}
\def\@arrayclassz{\ifcase \@lastchclass \@acolampacol \or
\@ampacol \or \or \or \@addamp \or
   \@acolampacol \or \@firstampfalse \@acol \fi
\edef\@preamble{\@preamble
  \ifcase \@chnum
     \hfil$\relax\arraymode\@sharp$\hfil
     \or $\relax\arraymode\@sharp$\hfil
     \or \hfil$\relax\arraymode\@sharp$\fi}}
\def\@array[#1]#2{\setbox\@arstrutbox=\hbox{\vrule
     height\arraystretch \ht\strutbox
     depth\arraystretch \dp\strutbox
     width\z@}\@mkpream{#2}\edef\@preamble{\halign
\noexpand\@halignto
\bgroup \tabskip\z@ \@arstrut \@preamble \tabskip\z@ \cr}%
\let\@startpbox\@@startpbox \let\@endpbox\@@endpbox
  \if #1t\vtop \else \if#1b\vbox \else \vcenter \fi\fi
  \bgroup \let\par\relax
  \let\@sharp##\let\protect\relax
  \@arrayskip\@preamble}
\def\eqnarray{\stepcounter{equation}%
              \let\@currentlabel=\theequation
              \global\@eqnswtrue
              \global\@eqcnt\z@
              \tabskip\@centering
              \let\\=\@eqncr
              $$%
 \halign to \displaywidth\bgroup
    \eqnumphantom\@eqnsel\hskip\@centering
    $\displaystyle \tabskip\z@ {##}$%
    \global\@eqcnt\@ne \hskip 2\arraycolsep
         $\displaystyle\arraymode{##}$\hfil
    \global\@eqcnt\tw@ \hskip 2\arraycolsep
         $\displaystyle\tabskip\z@{##}$\hfil
         \tabskip\@centering
    &{##}\tabskip\z@\cr}
\def\theequation{\thesection.\arabic{equation}}

\begin{document}
\begin{titlepage}
\setcounter{footnote}0
\begin{center}
\hfill PUPT-1722\\
\hfill ITEP/TH-42/97\\
\hfill hep-th/9709138\\
\vspace{0.5in}
{\LARGE\bf Seiberg-Witten Solution from Matrix Theory}\\
\bigskip
\bigskip
\bigskip
{\Large S.Gukov
\footnote{On leave from the Institute of Theoretical and
Experimental Physics, 117259 Bol. Cheremushkinskaya, 25,
Moscow, Russia}}\\
\bigskip
{\it Department of Physics, 
Joseph Henry Laboratories,\\
Princeton University\\
Princeton, NJ 08544, USA.\\
\bigskip
E-mail address: gukov@pupgg.princeton.edu}
\end{center}
\bigskip \bigskip

\begin{abstract}

As another evidence for the matrix Discrete Light
Cone formulation of M theory, we
show how general integrable Hamiltonian systems emerge from
BPS bound states of $k$ longitudinal fivebranes. Such
configurations preserve eight supercharges and by chain
of dualities can be related to the solution of $\CN=2$
four-dimensional gauge theories. Underlying Hitchin systems
on the bare spectral curve with $k$ singular points
arise from the 
Matrix theory compactification on the dual curve.

\end{abstract}

\end{titlepage}

\newpage
\setcounter{footnote}0
\section*{Introduction}
\setcounter{equation}{0}

Being a candidate for the nonperturbative formulation
of string theory
\cite{W1}, M theory plays a prominent role in the web of
dualities and understanding of nonperturbative dynamics in
various vacua. One of its notable applications is the
solution of $\CN=2$ SUSY gauge theories in four dimensions
\cite{W}. The Coulomb branch of the corresponding
field theory
comes as an effective theory of a
background brane configuration in string theory. From
the construction it is clear that integrability is
not inherent to four-dimensional physics, but seems to be
a characteristic feature of supersymmetric BPS state in string
theory. In fact, a spectral curve in \cite{W} appears as
a supersymmetric cycle which is a part of M-fivebrane
world-volume. Because the coupling constant of the effective
four-dimensional theory does not depend on the string coupling
constant, nonperturbative dynamics in $D=4$ may be extracted
by studying classical configurations in string theory.

There was no explicit formulation of M theory as
a quantum theory before the conjecture of \cite{BFSS}. 
In that
paper it was proposed to refer to $U(N)$ matrix quantum 
mechanics
of D-particles in the $N \ra \infty$ limit as exact 
description of M theory in the infinite momentum frame.
Then branes \cite{BFSS,GRT,BSS}, strings 
\cite{BS,DVV} and black holes \cite{DVV1,BFKS}
were realized in terms of the Matrix model;
their scattering
was shown as well to match corresponding string theory
amplitudes in many recent works. The matrix formulation
of M theory has already passed a number of tests via
compactification to various dimensions \cite{FHRS}
and in this line the Seiberg-Witten exact solution
\cite{SW} could be
another challenging consistency check for the Matrix model.

In the present work we make the first step in this direction
\footnote{Related efforts were also made in \cite{ZFC,P}}.
Surprisingly, the Matrix theory gives the clear physical answer
to the question: "Why integrability" from the first principles.
To begin with, let us emphasize what we exactly mean by 
{\it Matrix theory}. In order to accord to
$SU(N)$ gauge theory the solution require finite number $N$
of D0-brane charge. This is very similar to the 
compactification
of the Matrix theory on a three-torus which yields $\CN=4$ 
Yang-Mills
theory in $D=4$ that possesses exact S-duality group. 
This and
other arguments were accumulated
in \cite{S} to put the proposal \cite{BFSS}
further and to define finite $N$ formulation of 
the Matrix theory
in terms of the Discrete Light Cone Quantization (DLCQ).
Therefore, investigating finite $N$ dynamics, we give
another evidence for the DLCQ approach.

As was already mentioned above, supersymmetry is another
vital ingredient. In the next section we show that
 supersymmetry
arguments in the Matrix model restrict our attention
to longitudinal fivebrane backgrounds. In type IIA theory it
gives rise to bound state of D4-branes embedded in a
four-dimensional part $\CM$ of transverse space, D0-branes
and two mutually
orthogonal stacks of D2-branes also in $\CM$. The solution
is then defined by topological data only, e.g. genus of a 
spectral curve which
fixes the rank of gauge group and so on. Namely, in
our case a particular integrable system is selected out by
brane charges and manifold $\CM$.

Taking elliptic models as examples of Hitchin system \cite{H,H1},
in section 2 we investigate toroidal compactification of
the Matrix theory to 9 dimensions. This theory is equivalent
to type II string theory \cite{BS,FHRS},
and the longitudinal fivebrane background on $\CM$ implies
no D2-brane charges because of noncompact form of $\CM = R^2
\times T^2$ and finite $N$.
The bound state carry $N$ units of D0-brane charge
and $k$ units of $D4$-brane charge.
The fourbrane charges correspond to $k$ (dimensionally
reduced) "instantons" in the gauge theory on the dual
manifold $\CW$. The basic idea is that manifold $\CW$
does not include time variable of the ${\rm SYM}_{2+1}$
theory relevant to the Matrix theory compactification.
The "instantons" on $\CW$ are particles in $U(N)$
gauge theory on $R \times \CW = ({\rm time}) \times \CW$.
Therefore these {\it "instantons" on $\CW$
are point-like
states in $(2+1)$-dimensional
Super-Yang-Mills ( ${\rm SYM}_{2+1}$ )
theory on $R \times T^2$}.
We will show that incorporating them into
${\rm SYM}_{2+1}$
theory results in integrable Hitchin system on the
torus with $k$ punctures \cite{M,N}. Moreover, in the framework
of the Matrix theory all integrable systems and their spin 
generalizations
come on the same footing, reasserting the mightiness
 of the Matrix theory.
In the double scaling limit bare torus degenerates into
a sphere recovering the Seiberg-Witten exact solution.

Section 3 is devoted to string theory interpretation of the
results of section 2 and their relation to the solution of
$\CN=2$ gauge theories in four dimensions via M-fivebrane 
\cite{W}.
This clarifies the analogy between punctures on the bare spectral
curve and NS5-branes in brane configurations.

General properties of the longitudinal matrix fivebrane on
arbitrary $\CM$ in the form $\CM = \CC \times \CS$ 
and algebraically completely integrable
Hamiltonian systems behind such compactifications
are discussed in section 4.

We conclude in section 5 by some remarks on the rational
models, i.e. those relevant to theories with
fundamental matter.


\section{The Power of Supersymmetry}
\setcounter{equation}{0}

Ten-dimensional $U(N)$ Super-Yang-Mills theory 
reduced to $0+1$
dimensions has been proposed to describe M 
theory in the Discrete $N$
Light Cone formalism \cite{BFSS,S}. 
Eleven-dimensional action written
in terms of $N \times N$ Hermitean 
matrices $X$ and $\Psi$:
\be
S= \Tr \int dt  \BL {1 \over 2R} (D_0 X^i)^2 - 
{1 \over 4} R \[ X^i , X^j \]^2 - 
{\bar \Psi} D_0 \Psi -
R {\bar \Psi} \Ga^i \[ X_i , \Psi \] \BR
\label{11d}
\ee
depends on the Plank length $l_p$ and the radius of the
compact eleventh direction $R$.

According to \cite{BFSS}, compactification of the original
M theory on some manifold $\CM$ opens up extra dimensions
in Yang-Mills theory on the manifold $\CW$ of the inverse size
\footnote{$\CM$ does not include the 11th direction which will
be assumed large for the rest of the paper.}.
For example, $L_1, \ldots , L_d$ toroidal compactification on
$\CM = T^d$ leads to the $(d+1)$-dimensional Yang-Mills
theory on the dual torus $\CW = {\tilde T}^d$ with
sides \footnote{We omit factors of $2 \pi$.} :
\be
\Sig_i = {l_p^3 \over R L_i}
\label{t}
\ee

${\rm SYM}_{d+1}$ coupling constant \cite{FHRS}
\be
g^2_{YM} = {R^3 \over l_p^6} \prod_i 
\BL {l_p^3 \over R L_i} \BR
\label{g}
\ee
goes to infinity when the compact 
manifold $\CM$ shrinks to zero size.
In this limit one has to replace 
$N \times N$ matrices $X$ by covariant
derivatives:
\be
X_{\mu} \ra -i D_{\mu} = 
-i ( {\di \over \di x_{\mu}} + A_{\mu})
\label{xd}
\ee
with respect to $G=U(N)$-valued gauge connection $A_{\mu}$,
so that the corresponding action takes the following form:
\be
S=  {1 \over g^2_{YM}} \Tr \int_{R \times \CM} dt d^d x 
\BL {1 \over 4} F_{\mu \nu} F^{\mu \nu}
- \half {\bar \Psi} \Ga^{\mu} \[ X_{\mu} , \Psi \] \BR
\label{act}
\ee
where $F_{\mu \nu} = \[ X_{\mu}, X_{\nu} \]$ 
with the appropriate
replacement (\ref{xd}) of compact $X_{\mu}$. Here and in the
following the indices $\mu$, $\nu$ run from $0$ to $9$.

Action (\ref{act}) enjoys 16 dynamical SUSY transformations:
\be
\de X_{\mu} = i {\bar \ep} \Ga_{\mu} \Psi \qquad \mu = 0,1,
\ldots , 9 \\
\de \Psi = (D_0 X_i) \Ga^{0i} \ep + {i \over 2}
\[ X_i, X_j \] \Ga^{ij} \ep \qquad i,j = 1,2, \ldots , 9
\label{dsusy}
\ee
and 16 kinematical supersymmetries:
\be
\de X_{\mu} = 0 \qquad
\de \Psi = {\tilde \ep}
\label{ksusy}
\ee
which have nonlinear realization and are never preserved by
themselves. For this reason in the sequel we will search for
BPS states preserving half of the dynamical SUSY only, i.e.
eight real supercharges.

Antisymmetric in $\mu, \nu$ matrix $F_{\mu \nu}$ can always
be brought to Jordan form:
\be
F_{\mu \nu} = \pmatrix{
 0 & F_{01} & 0 & 0 & 0 & 0 & 0 & 0 & 0 & 0 \cr
 - F_{01} & 0 & 0 & 0 & 0 & 0 & 0 & 0 & 0 & 0 \cr
 0 & 0 & 0 & F_{23} & 0 & 0 & 0 & 0 & 0 & 0  \cr
 0 & 0 & - F_{23} & 0 & 0 & 0 & 0 &  0 & 0 & 0 \cr
 0 & 0 & 0 & 0 & 0 & F_{45} & 0 & 0 &  0 & 0 \cr
 0 & 0 & 0 & 0 & - F_{45} & 0 & 0 & 0 & 0 & 0 \cr
 0 & 0 & 0 & 0 & 0 & 0 & 0 & F_{67} & 0 &  0 \cr
 0 & 0 & 0 & 0 & 0 & 0 & -F_{67} & 0 & 0 & 0  \cr
 0 & 0 & 0 & 0 & 0 & 0 & 0 & 0 & 0 & F_{89}  \cr
 0 & 0 & 0 & 0 & 0 & 0 & 0 & 0 & -F_{89} & 0 }
\label{jord}
\ee

The matrix $F$ corresponding to
Dp-brane BPS state preserving 
${1 \over 2^{({p \over 2} -1)}}$ fraction
of the dynamical SUSY (\ref{dsusy}) 
satisfies the following conditions \cite{HLW}:
\be
F_{12} - \sum_{i=2}^{{p \over 2}} 
\xi_i F_{(2i-1)(2i)} = 0 \qquad
\Ga_{12} \Ga^{(2i-1)(2i)} \ep = \xi_i \ep
\label{fsusy}
\ee

Therefore, eight real supercharges are preserved for a BPS
background with, say $F_{45} = F_{67}$ and all the
other fields in
(\ref{act}) put to zero. This background corresponds to the
marginal bound state of a $D(p=4)$-brane  with $N$ D0-branes
and two equal sets of D2-branes \cite{HLW,T2}.
The total Dp-brane topological charge of the field 
configuration is \cite{GRT,BSS}:
\be
Q_p \propto \int \Tr F^{\wedge ({p \over 2})} =  
\int \Tr F \wedge \ldots \wedge F
\label{charge}
\ee
Note that like in \cite{W} supersymmetry arguments lead us
to the longitudinal fivebrane background, but
now in the very different way.

Because all the background
fields in the directions other than $x^4$, 
$x^5$, $x^6$ and $x^7$
must vanish, we consider $k$ longitudinal fivebranes
on a four-manifold $\CM =  (x^4, 
x^5, x^6, x^7)$. From the Yang-Mills
side it means charge
$k$ instanton solution in $U(N)$ 
gauge theory on $\CW$ \cite{T2}:
\be
k = {1 \over 8 \pi^2} \int \Tr F \wedge F
\label{k}
\ee
which has a moduli space of self-dual (or anti-self-dual for
negative $k$) field strengths ${\tilde F} = \pm F$
\footnote{We always imply covariant derivative (\ref{xd})
for compact $X$.}:
\begin{equation} \ \ \ \left\{
\begin{array}{c}
F_{45} = \[ X_4, X_5 \] = \[ X_6 , X_7 \] = F_{67} \cr
F_{57} = \[ X_5, X_7 \] = \[ X_6 , X_4 \] = F_{64} \cr
F_{74} = \[ X_7, X_4 \] = \[ X_6 , X_5 \] = F_{65}
\end{array}\right.
\label{sdual}
\end{equation}

Self-duality of the field $F$ on $\CW$ is the key point in our
construction and we will make use of it in the spirit
of \cite{H1} and \cite{BJSV}. The crux of the matter
is that from the Matrix theory compactification on $\CM$ we get
$(4+1)$-dimensional $U(N)$ gauge theory on
$R \times \CW = ({\rm time}) \times \CW$
with $k$ four-dimensional "instantons" on $\CM$
which does not include time! From the
${\rm SYM}_{4+1}$ point of 
view {\it these "instantons" look like $k$
particles on $R \times \CW$}. We will
use this trick in the next section to make a reduction
with respect to a subgroup
and impose a moment map condition. The important
role of the self-dual field was also discussed in \cite{BDL}
for the equivalent (by T-duality in $x^6$, $x^7$ directions)
configuration of two-branes intersecting at angles.

Let us stress here the essential role of supersymmetry.
Indeed, all the arguments we made so far were based on 
preserving a quarter of SUSY in the Matrix theory.
In the next sections we will show 
that specifying the manifold $\CM$
we determine the system completely which has a natural 
integrable structure in the holomorphic sense \cite{H}.
In what follows we will consider only compactifications
of the Matrix theory on a direct product of
two complex surfaces $\CM = \CC\times \CS$
parametrized by (local) holomorphic coordinates:
\be
w=x^4+i x^5 \in \CC
\qquad {\rm and} \qquad
z=x^6+i x^7 \in \CS
\label{coord}
\ee
and let us take $\CC = R^2$ and $\CS = T^2$ as
the first example.


\section{Elliptic Models}
\setcounter{equation}{0}

Elliptic integrable models usually arise in $(2+1)$- and
$(1+1)$-dimensional topological gauge theories on the same
torus \cite{DW,GN,GN1,GM}. This is a strong motivation to study
toroidal compactification of the Matrix theory which gives
rise to Super-Yang-Mills in $2+1$ dimensions \cite{FHRS,BFSS1}.
By this reason in the present 
section we examine a compactification
of $k$ longitudinal fivebranes on $\CM = \CC \times \CS =
R^2 \times T^2$, where the torus:
\be
y^2=(x-e_1)(x-e_2)(x-e_3) \nn\\
x=\wp (z) \qquad y=\wp'(z) \qquad dz = {dx \over y}
\label{torus}
\ee
has sides $L_6$ and $L_7$, and $\wp (z)$ is the double periodic
Weierstrass $\wp$-function with periods 
$1$ and $\tau = {\th \over 2 \pi}+ {4 \pi i \over g^2}$.
${\rm SYM}_{2+1}$ theory is defined 
on the dual torus ${\tilde \CS}$ with
the same complex structure but with the inverse sides
$\Sig_6 = {l_p^3 \over R L_6}$ and
$\Sig_7 = {l_p^3 \over R L_7}$.

In these notations self-duality equations (\ref{sdual})
for $F_{\mu \nu}$ take the form:
\be
F = {i \over 2} \[ \phi , {\bar \phi} \] \nn \\
{\bar \CD} \phi = 0
\label{self}
\ee
where
$\CD = D_6 + i D_7$, $F = {i \over 2} \[ \CD, {\bar \CD} \]$,
$\phi = X^4 + i X^5$ and the gauge connection $A = A^6 + i A^7$
take values in the complexified Lie algebra $G_C = U_C (N)$ for
the case at hand.

Remarkably, the second equation in (\ref{self}) is nothing but a
Gauss law in the $(2+1)$-dimensional gauge theory on $R \times
{\tilde \CS}$.
This can be easily seen by introducing a source term
in the action (\ref{act}):
\be
\delta S_{{\rm source}} = 
\Tr \int_{R \times T^2} dt d^2 z A_0 J_0
\delta (z)
\label{sact}
\ee

Concentrating on holomorphic bundles over the torus,
the relevant part of the action (\ref{act}) looks like
\cite{GM}:
\be
S = {1 \over g_{YM}^2} \Tr \int_{R \times T^2} dt d^2 z
\[ - F_{0{\bar z}} \Phi + 
\half \Phi^2 + A_0 J_0 \delta (z) + \ldots \] = \nn \\
=  {1 \over g_{YM}^2} \Tr \int_{R \times T^2} dt d^2 z
\[ A_0 \( - {\bar \CD} \Phi  + 
J_0 \delta (z)\) + {1 \over 2} \Phi^2 + \ldots \]
\label{ract}
\ee
where we have introduced 
an auxiliary field $\Phi = F_{0 {\bar z}}$.

But one shall not forget about $k$ "instantons" on $\CW$ or,
equivalently, about $k$ particles on $R \times \CW$.
The "instantons" we discuss here resemble actual
four-dimensional instantons only when $\CM$ is
compact, i.e. $\CW$ is four-dimensional. In the
elliptic case of this section an "instanton"
means a self-dual solution to the equations
(\ref{sdual}) with non-zero charge (\ref{k}) in two
dimensions. For this reason we distinguish it
by quotation-marks from the four-dimensional
instanton.

Therefore, implementing $k$ "instantons" means taking $k$ source
terms (\ref{sact}) that modify the Gauss law (\ref{self})
to \cite{N,G,MMM}:
\be
{\bar \di} \Phi_{ij}+(a_i - a_j) \Phi_{ij} = \sum_{\a =1}^k
J_{ij}^{(\a)} \de (z -z_{\a})
\label{BPS}
\ee
where we used the
residual gauge symmetry on the torus to take $A$ in the
diagonal form $A= {\rm diag} (a_1, \ldots, a_N)$.
Matrices $J_{ij}^{(\a)}$ determine the orientation of
$(2+1)$-dimensional particles in the color group.
Physically, taking $A_0= 0$ gauge, 
one must bear in mind the Gauss law (\ref{BPS})
which turns out to be a moment map condition!

Let us
outline the entire chain of points we followed to
obtain the moment map equation (\ref{BPS}) from
the $D4-D0$ bound state in the Matrix theory:

\be
\begin{array}{c}
\mbox{\underline{Supersymmetric 
vacuum}} \cr
\downarrow \cr
\mbox{\underline{Longitudinal 
fivebrane on $\CM$ }} \cr
\downarrow \cr
\mbox{\underline{Self-dual field on
$\CW$ }} \cr
\downarrow \cr
\mbox{\underline{$U(N)$ "instanton" on
$\CW$ }} \cr
\downarrow \cr
\mbox{\underline{A particle
in ${\rm SYM}_{2+1}$ on
$R \times \CW$ }} \cr
\downarrow \cr
\mbox{\underline{The moment map condition}}
\end{array}
\ee

In the language of integrable system each source (\ref{sact})
represents a marked point on the bare spectral curve
${\tilde \CS}$ of the corresponding Hitchin system
\cite{M,N,G,DW}. It means that the
number of longitudinal matrix fivebranes $k$ is
exactly
the number of singular points on ${\tilde \CS}$. Moreover,
in the Matrix program integrable data appear
in a clear physical way, e.g. stable pair $(V , \Phi)$
of Hitchin system is defined by $G = U_c (N)$ vector bundle
and the adjoint field $\Phi$ of $U(N)$
${\rm SYM}_{2+1}$ compactification.

Nontrivial solution to the moment map equation (\ref{BPS}):
{\footnotesize
\be
\Phi_{ij} (z) = \de_{ij} \BL p_i + \sum_{\a} J_{ij}^{(\a)} \di
\log \th (z-z_{\a} \vert \t) \BR + (1 - \de_{ij})
e^{a_{ij}(z- {\bar z})} \sum_{\a} J_{ij}^{(\a)}
\frac{\th (z-z_{\a}+{Im \t \over \pi} a_{ij}) \th' (0)}
{\th (z-z_{\a}) \th ({Im \t \over \pi} a_{ij})}
\label{cal}
\ee}
is nothing but the Lax operator
for general elliptic Gaudin system \cite{N},
where we have denoted $a_{ij} = a_i - a_j$
for brevity.
A gauge transformation:
\be
\Phi_{ij} \ra (U^{-1} \Phi U)_{ij} (z) \qquad U_{ij}
= e^{a_{ij} {\bar z}}
\label{gtr}
\ee
takes it to the holomorphic form:
{\footnotesize
\be
\Phi_{ij} (z) = \de_{ij} \BL p_i + \sum_{\a} J_{ij}^{(\a)} \di
\log \th (z-z_{\a} \vert \t) \BR + (1 - \de_{ij})
e^{a_{ij} z} \sum_{\a} J_{ij}^{(\a)}
\frac{\th (z-z_{\a}+{Im \t \over \pi} a_{ij}) \th' (0)}
{\th (z-z_{\a}) \th ({Im \t \over \pi} a_{ij})}
\label{gaudin}
\ee}

Introducing the spectral curve:
\be
P (\la,z) = 
\det_{N \times N} (\la \de_{ij} - \Phi_{ij} (z)) = 0
\label{curve}
\ee
and the meromorphic 1-differential $dS$ on it:
\be
dS = \la dz \qquad
{\di dS \over \di h_j} = {\rm holomorphic ~differential}
\label{ds}
\ee
one can easily find the solutions of string vacua or $\CN=2$
field theories. For instance, masses of BPS multiplets
are defined by periods of $dS$:
\be
A_i = \int_{\a_i} dS \qquad A^D_i = \int_{\b_i} dS
\label{aad}
\ee
over the basic cycles
$\a_i \circ \b_j = \de_{ij}$ on the spectral
surface (\ref{curve}).

Somewhat miraculously,
unlike (\ref{k}), two-brane charge
$\int \Tr F = \int \Tr \[ X^4, X^5 \]$ vanishes
identically because of two noncompact directions $x^4$
and $x^5$ and finite $N$. Hence the extra constraint on
matrices $J_{ij}^{(\a)}$ \cite{N}:
\be
\sum_{\a =1}^k J_{ii}^{(\a)} =0
\label{trace}
\ee
is automatically satisfied in the realm of the Matrix theory.
The absence of D2-branes in the system will also prove essential 
in the next section where we study the relation of this system to
the brane construction of \cite{W}. If there were
nonzero D2-brane charges, it would be not possible to map
the matrix brane configuration 
to that of \cite{W} where it was used
to derive the solutions of $\CN=2$ gauge theories in four
dimensions. Present approach naturally generalizes elliptic
models of \cite{W} to spin degrees of freedom. Namely,
the solution (\ref{gaudin}) in the most general form
corresponds to the elliptic Gaudin model.

For the sake of simplicity, let us take an illustrative example
of elliptic Calogero-Moser model. This system describes
$N$ pair-wise interacting particles on the torus
with the Hamiltonian:
\be
H=h_2 = \sum_i {p_i^2 \over 2} + {m^2 \over 8} \sum_{i < j}
\wp (x_i -x_j)
\label{hcal}
\ee
Lax operator (\ref{gaudin}) for this model
follows from the solution of
the moment map equation (\ref{BPS}) with the only marked
point where the residue is $J_{ij} = m (1 - \de_{ij})$
which means taking a brane
background with $N$ D0-branes and $k=1$.

In the double-scaling limit
\be
m^2 \ra \infty \qquad q=e^{i \pi \t} \ra 0
\label{2sc}
\ee
so that $\La^N \propto m^N q$ 
remains finite, the system goes to the 
periodic Toda chain which is behind the integrability of
$\CN=2$ supersymmetric $SU(N)$ pure gauge theory in four
dimensions \cite{GKMMM}. In the brane construction
\cite{W} it also occurs as degeneration of the elliptic model
that gives a hint for the equivalence of the bare spectral curve
(\ref{torus})
and space-time torus in \cite{W}. The analogy will be
corroborated in the next section.

$N$-periodic Toda chain describes one-dimensional system of
$N$ particles on a circle with exponential interaction
of the nearest neighbours. In this limit the spectral
curve (\ref{curve}) degenerates into double covering of
sphere and takes a very simple form:
\be
\La^N \cosh (z) = 2 P_{(N)} (\la)
\label{tcurve}
\ee
that coincide with that of the Seiberg-Witten solution
of $\CN=2$ four-dimensional gauge theory \cite{SW,GKMMM}.

Another reward of the Matrix theory is that the prepotential
and vacuum expectation values of $\CN=2$ gauge theory also
come into the story. Thus, order parameters on the Coulomb
phase are functionally independent, Poisson-commuting
Hamiltonians of the integrable dynamics \cite{DW,GKMMM}:
\be
h_j = {1 \over j} \langle \phi^j \rangle \qquad
h_j \approx {1 \over j} \sum_{i=1}^N p_i^j + \ldots
\qquad
\langle \phi^j \rangle \approx 
\sum_{i=1}^N A_i^j + \ldots
\label{h}
\ee
The number of Hamiltonians $h_j$ is the same as the genus
of the spectral curve (\ref{curve}) $g = Nk -
{k(k+1) \over 2} +1$ from the Riemann-Roch theorem.

The homogeneity of differential (\ref{ds}) allows one
to express the prepotential $\CF$ via Hamiltonians (\ref{h}).
For the particular choice of $SU(2)$ gauge group, $\CF$ is a
solution to the equation \cite{MAT}:
\be
a{\di \CF \over \di a} - 2 \CF = {2 i \over \pi} H
\label{su2}
\ee


\section{M-branes versus Matrix branes}
\setcounter{equation}{0}

In this section we prove that the brane
configuration of \cite{W} corresponding to the $\CN=2$
four-dimensional SYM theory with adjoint hypermultiplet
represents the same
background as the Matrix theory setup. For
this purpose let us briefly remind the brane picture
\cite{W}.

Classically we take $N$ 
D4-branes with $x^0, x^1, x^2, x^3, x^6$
world-volume coordinates to get $U(N)$ Yang-Mills gauge theory
with 16 real supercharges. In order to have $\CN=4$
four-dimensional
low-energy field theory, one has to compactify $x^6$ direction
on a circle of  radius $Y_6 = {g_{st} l_s \over g^2}$.
The fourbranes are all at the same position
along the $x^7, x^8, x^9$ directions and have different
$v = x^4 + i x^5$ coordinates.
To break supersymmetry further by a half one
introduces a fivebrane with the world-volume in
$x^0, x^1, x^2, x^3, x^4,
x^5$. The space-time metric allows
a nontrivial ${\bf C}$-bundle over $S^1$:
\be
x^6 \ra x^6 + Y_6 \qquad v \ra v + m \qquad x^{11} \ra x^{11}
+ Y_{11} \th
\label{6bundle}
\ee
with arbitrary parameter $m$ corresponding to the bare mass
of the adjoint $\CN=2$ hypermultiplet that softly breaks
$\CN=4$ supersymmetry. Varying $m$ we interpolate between
$\CN=4$ $SU(N)$ SYM theory ($m=0$) and pure gauge $\CN=2$
theory (in the double scaling limit (\ref{2sc})).
Now let us turn to the Matrix picture.

As it was explained in the previous section, type IIA theory
description of theories with 8 real 
supercharges comes from the
$4-2-2-0$ bound state. We choose the 
world-volumes of the branes
to span $x^4, x^5, x^6, x^7$ for the D4-branes,
$x^4$ and $x^5$ for the first set of D2-branes,
$x^6$ and $x^7$ for the second set of D2-branes
and D0 branes are allowed to move in all of
the $x^4, x^5, x^6, x^7$
directions. If we restrict ourselves to the case of $U(N)$
gauge theory with adjoint matter, we have to consider
the only D4-brane with $N$ units of D0-brane charge
without twobranes. In the language of 
the Matrix theory it means taking a longitudinal
fivebrane in the DLCQ formulation of $U(N)$ matrix model at
finite $N$. Moreover, for this particular case (of elliptic
Calogero system) $x^6$ and $x^7$ must compound the bare
torus (\ref{torus}) with modular parameter $\t$.

The equivalence of the two brane configurations comes from the
chain of dualities under which a fivebrane goes into matrix
longitudinal fivebrane, and each 
D4-brane gets mapped into a D0-brane.
If we start from the matrix picture of $D4-D0$ bound state:
\begin{equation} \ \ \ \left\{
\begin{array}{c|cccccccccc}
      & 0& 1& 2& 3& 4& 5& 6& 7& 8& 9\cr
D4    & +& -& -& -& +& +& +& +& -& -\cr
D0    & +& -& -& -& -& -& -& -& -& -
\end{array}\right.
\label{tabl0}
\end{equation}
the sequence is the following:
\begin{itemize}

\item{T-duality along $x^6$ and $x^7$}

It takes the $T^2$ to the dual 
torus with inverse size according
to (\ref{t}), but does not change its complex
structure. Therefore, sides of the torus become $\Sig_6$
and $\Sig_7$ of the Yang-Mills theory.
And the brane configuration changes to:
\begin{equation} \ \ \ \left\{
\begin{array}{c|cccccccccc}
      & 0& 1& 2& 3& 4& 5& 6& 7& 8& 9\cr
D2    & +& -& -& -& +& +& -& -& -& -\cr
D2    & +& -& -& -& -& -& +& +& -& -
\end{array}\right.
\label{tabl1}
\end{equation}
where a "plus" stands for the extended direction of a brane
and a "minus" corresponds to its fixed position.

\item{T-duality along $x^1, x^2, x^3$}

This duality makes two sets of D5-branes 
in type IIB string theory with four common
directions $x^0, x^1, x^2$ and $x^3$ which will be the
space-time of the effective four-dimensional theory:
\begin{equation} \ \ \ \left\{
\begin{array}{c|cccccccccc}
      & 0& 1& 2& 3& 4& 5& 6& 7& 8& 9\cr
D5    & +& +& +& +& +& +& -& -& -& -\cr
D5    & +& +& +& +& -& -& +& +& -& -
\end{array}\right.
\label{tabl2}
\end{equation}

Finally, to get the brane configuration of \cite{W}, we
recall the eleventh dimension and perform

\item{$7 \leftrightarrow 11$ Flip}

It does not change anything from the M theory point of view
because of the full 11-dimensional Lorentz invariance (but,
of course, modifies the coupling 
constant $g_{st}$ of type IIA string theory).
We come to the single fivebrane:

\begin{equation} \ \ \ \left\{
\begin{array}{c|ccccccccccc}
      & 0& 1& 2& 3& 4& 5& 6& 7& 8& 9& 11\cr
NS5   & +& +& +& +& +& +& -& -& -& -& -\cr
D4    & +& +& +& +& -& -& +& -& -& -& +
\end{array}\right.
\label{tabl3}
\end{equation}
that has $R^4
\times \Sigma$ world-volume, where $\Sigma$ is 
the spectral curve (\ref{curve}) of the
underlying integrable system holomorphically embedded into
$R^2 \times T^2$ where the torus ${\tilde \CS}$
has periods $1$ and $\t$ and sides $\Sig_6$ and $\Sig_7$.
The classical picture of $N$ $D4$-branes suspended in
the background of NS5-brane is restored in the type IIA
limit when radius of the 11th dimension goes to zero.

\end{itemize}

We have established the perfect agreement between the Matrix
solution via longitudinal fivebrane and the standard M theory
approach \cite{W}. One can apply this (or reverse) chain of
dualities to relate any brane constructions of field theories
in the Coulomb phase to their matrix counterparts.


\section{General $\CM$}
\setcounter{equation}{0}

In this section we make several remarks on the Matrix theory
compactification on four-manifold $\CM$ in the most general
form $\CM = \CC \times \CS$. The most straightforward one
is the extension of the analysis in the previous sections to
arbitrary genus $g > 1$ compact Riemann surface ${\tilde \CS}$
while $\CC$ is still a complex plane. 
This is exactly the subject
of \cite{H} where it was shown that 
even without singular points
on ${\tilde \CS}$ we end up with an algebraically
completely integrable Hamiltonian system.
In fact, holomorphic vector bundle $V$ of rank $N$ over the
compact smooth Riemann surface $\tilde \CS$ and its section
$\Phi$ arise from $U(N)$ ${\rm SYM}_{2+1}$ theory on
$R \times {\tilde \CS}$ in a very natural way. By the
vanishing theorem \cite{H1}, the pair $(V , \Phi)$
arising from a solution of the self-duality equations
(\ref{self}) is necessarily stable. Cotangent bundle
$T^{\*} M_V$ of the moduli space $M_V$ of this stable
bundle $V$ is Hitchin system, because the number
of Poisson-commuting Hamiltonians $h_j$ is exactly
the same as the dimension of the moduli space
${\rm dim} (M_V) = N^2 (g-1) +1$ \cite{H,M}.

Another possible compactification is on $\CM = \CC \times \CS$
where the size of the compact manifold
$\CC$ goes to infinity and fivebrane charge $k$ is turned
off. On the Yang-Mills
side it looks like compactification of
$\CN=4$ (instead of $\CN=2$) four-dimensional
$U(N)$ Yang-Mills theory without 
"source" on $\CW = {\tilde \CC}
\times {\tilde \CS}$ where $V_{\tilde \CC} \ra 0$.
Similar reduction of slightly different (twisted) $\CN=4$
theories was shown to yield supersymmetric 2D sigma-model on
hyperKahler Hitchin space (namely, compactified cotangent
bundle $T^{\*} M_V$ to the moduli space $M_V$ of flat
connections) \cite{BJSV}.

Finally, one can consider compactification on four-manifold
$\CM$ with a shrinking three-cycle in it. It results in
$\CN=2$ supersymmetric four-dimensional $U(N)$ Yang-Mills
theory in strong coupling regime. The same reasoning
as in \cite{FHRS} explains that it probes infrared
behaviour of the theory where we again come to the
usual Seiberg-Witten story \cite{SW}.

It seems possible to continue the list of integrable
string vacua by introducing D2-branes and extending
to various $\CM$. The arguments of this section and
detailed study of elliptic models in section 2 strongly
suggest integrability of arbitrary vacuum in the Matrix
theory preserving eight supercharges. We hope that
this assumption will be illuminated in future.
Note, that using the results of section 3 one can
conjecture that the fivebrane of \cite{W} lives on the {\it dual}
space $\CC \times {\tilde \CS}$, i.e. spectral curve of the
solution must be holomorphically embedded in
$\CC \times {\tilde \CS}$.


\section{Discussion on Rational Models}
\setcounter{equation}{0}

Applying the basic idea of section 4 to toroidal
compactification of the longitudinal fivebrane in
the DLCQ formulation
of the Matrix theory, we derived in section 2 the
elliptic Hamiltonian
systems of Hitchin type. Had we been aware of integrability
in the related brane background \cite{W}, we would have found
it in the dynamics of $4-2-2-0$ bound state.
This approach naturally extends known integrable string
theory vacua to spin degrees of freedom of the integrable
counterparts, but the original Seiberg-Witten solution
\cite{SW} arise only in the double scaling limit (\ref{2sc})
as the torus ${\tilde \CS}$ degenerates into a sphere. This
poses a problem how to obtain rational models also in
a straightforward way. This is a very interesting
subject for future work, because rational theories, i.e.
$\CN=2$ four-dimensional gauge theories with fundamental
matter generally corresponding to spin magnets \cite{GGM},
are solvable due to the pivotal concept of R-matrix
which is not manifest in the present approach.

Now we sketch an indirect argument for the equivalence
of integrable inhomogeneous spin chains and certain
compactifications of Heterotic Matrix theory
\cite{BM,HOR}. Completely
different way to break supersymmetry (from what we have
used in the main part of the text) is to study a membrane
compactification on the orbifold $I \times S^1$. Corresponding
$(2+1)$-dimensional Yang-Mills theory defined on the
dual manifold $S^1 \times {\tilde \CS} = I \times T^2$
has the desired SUSY.

The first sign of the Ruijsenaars system can be seen already
at this stage from the D2-brane defining equation
\cite{BFSS,GRT,BSS}:
\be
UVU^{-1}V^{-1} = Z
\label{membr}
\ee
This monodromy over the torus in the Chern-Simons theory
gives rise \cite{GN1} to the spectrum of the
Ruijsenaars model.

The theory described above corresponds to compactification
of Heterotic Matrix theory where supersymmetry and gauge
symmetry anomaly cancellation
arguments require to include a Chern-Simons term:
\be
S_{CS} = {\kappa \over 2} \Tr \int ( AdA + {2i \over 3} A^3)
\label{cs}
\ee
to the $(2+1)$-dimensional Yang-Mills action \cite{KR,HOR}.
It modifies equations of motion of 
the Matrix model to \cite{ZFC}:
\be
{\di^2 X^{\mu} \over \di t^2} = - \kappa^2 X^{\mu} -
{3 \kappa \over 2} \eps^{\mu \nu \eta}
\[ X^{\nu}, X^{\eta} \] - \[X^{\nu} \[ X^{\mu}, X^{\nu} \] \]
\label{zfc}
\ee
which have common solutions with the first-order equation:
\be
{\di X^{\mu} \over \di t} = - i \kappa X^{\mu} +
{i \over 2} \eps^{\mu \nu \eta}
\[ X^{\nu}, X^{\eta} \]
\label{zbps}
\ee

They are nothing but the
Nahm equations for $SU(2)$ monopole of charge $N$ under
the appropriate change of variables \cite{H2}:
\be
{\di T_{i} \over \di t} = 
{1 \over 2} \eps^{ijk}
\[ T_j, T_k \]
\label{nahm}
\ee
The latter
is equivalent to the inhomogeneous XXX $Sl(N)$ spin chain
which governs the dynamics of $SU(2)^{N-1}$ gauge theory with
fundamental matter \cite{GGM}.

The shortcut to this lengthy solution of models corresponding
to theories with fundamental matter as well as other aspects
of Heterotic Matrix theory compactification make up the
subject of future investigation.



\section*{Acknowledgments}

I am grateful to A.Gorsky,
I.~Klebanov, A.~Mikhailov, A.~Mironov,
A.~Morozov, N.~Nekrasov and W.~Taylor
for enlightening discussions.

The work was supported in part by Merit Fellowship
in Natural Sciences and Mathematics and grant RFBR-96-15-96939.


\end{document}